\documentclass[aps,prl,twocolumn,showpacs,amsmath,amssymb,amsfonts,nofootinbib,long]{revtex4}
\usepackage{epsfig,latexsym,bm}

\newcommand\GeV{\mbox{GeV}}
\newcommand\MeV{\mbox{MeV}}

\newcommand\K{\mbox{K}}

\begin{document}

\title{Helium-4 Synthesis in an Anisotropic Universe}

\author{Leonardo Campanelli$^{1}$}
\email{leonardo.campanelli@ba.infn.it}

\affiliation{$^1$Dipartimento di Fisica, Universit\`{a} di Bari, I-70126 Bari, Italy}

\date{December, 2011}


\begin{abstract}
\begin{center}
{\bf Abstract}
\end{center}
We calculate the $^4$He abundance in a universe of Bianchi type I whose cosmic anisotropy is
dynamically generated by a fluid with anisotropic equation of state. Requiring that the relative
variation of mass fraction of $^4$He is less than $4\%$ with respect to the standard isotropic
case to be consistent with astrophysical data, we constrain the parameter of cosmic anisotropy,
the shear $\Sigma$, as $|\Sigma(T_f)| \lesssim 0.4$, where $T_f$ is the freeze-out temperature
of the weak interactions that interconvert neutrons and protons. Anisotropic fluids, whose energy
density is subdominant with respect to the energy content of the Universe during inflation and
radiation era, generate much smaller shears at the time of freeze-out and then do not appreciably
affect the standard $^4$He production. This is the case of anisotropic dark energy, and of a
uniform magnetic field with energy density much smaller than about $1.25$ times the energy
density of neutrinos.
\end{abstract}


\pacs{98.80.Jk, 98.80.Ft}

\maketitle


\section{\normalsize{I. Introduction}}
\renewcommand{\thesection}{\arabic{section}}

The high level of isotropy of the cosmic microwave background (CMB)
radiation is the most convincing justification of the Cosmological Principle:
the Universe is homogeneous and isotropic at large cosmological scales~\cite{Weinberg}.
However, tiny deviations from perfect isotropy are not excluded by present CMB data.
Indeed, a particular anisotropic cosmological model of Bianchi type I, known as
{\it ellipsoidal universe}~\cite{ellipsoidal1,ellipsoidal2},
can better match CMB data and solve the so-called ``quadrupole problem,''
namely the lack of CMB power detected on large angular scales.

Various mechanisms could give rise to an ellipsoidal universe, such as
a uniform cosmological magnetic field~\cite{Barrow1,ellipsoidal1,ellipsoidal2},
topological defects (e.g. cosmic stings, domain walls)~\cite{Barrow1},
or a dark energy fluid with anisotropic equation of state~\cite{Barrow1,Koivisto-Mota}.
Independently on the nature of the mechanism, however,
a modification of the standard picture of primordial nucleosynthesis can occur
if universe anisotropization takes place during the early Universe~\cite{R1,R2,R3,R4}.

The aim of this paper is, indeed, to constrain the level of
cosmic anisotropy, so as to be consistent with observational bounds on
primeval $^4$He abundance.


\section{\normalsize{II. Ellipsoidal Universe}}
\renewcommand{\thesection}{\arabic{section}}

The ellipsoidal
universe~\cite{ellipsoidal1,ellipsoidal2,ellipsoidal3,ellipsoidal4,ellipsoidal5,ellipsoidal6,ellipsoidal7,ellipsoidal8}
is a cosmological model described by the Bianchi I line element~\cite{Taub}
\begin{equation}
\label{metric}
ds^2 = dt^2 - a^2(t) (dx^2 + dy^2) - b^2(t) \, dz^2
\end{equation}
with two scale factors, $a$ and $b$, normalized as $a =
b= 1$ at the present cosmic time. Cosmic anisotropy
is quantified by the shear
\begin{equation}
\label{Sheardef} \Sigma = \frac{H_a - H_b}{2H_a + H_b} \, ,
\end{equation}
with $H_a = \dot{a}/a$ and $H_b = \dot{b}/b$, while
$H = \dot{A}/A = (2H_a + H_b)/3$ and $A = (a^2b)^{1/3}$ play
the usual role of Hubble and expansion parameters, respectively.
(Here and in the following a dot indicates a derivative with respect
to cosmic time $t$).

Anisotropy of the Universe is not assumed a priori but dynamically
generated by an anisotropic fluid (A) with two equations of state:
$w^{\|}_A = p^{\|}_A/\rho_A$ and $w^{\! \bot}_A = p^{\bot}_A/\rho_A$,
where $p^{\|}_A$ and $p^{\bot}_A$ are respectively the pressures along the
$x$ ($y$) and $z$ directions, and $\rho_A$ the energy density. The source of
cosmic anisotropy is then parameterized by the skewness
$\delta_A = w^{\|}_A - w^{\! \bot}_A$, while $w_A = (2w^{\|}_A + w^{\! \bot}_A)/3$
represents the usual equation of state parameter.

Friedmann equation in ellipsoidal universe takes the
form~\cite{ellipsoidal6,ellipsoidal7}
\begin{eqnarray}
\label{H}
(1-\Sigma^2) H^2 = \frac{8\pi G}{3} \, (\rho + \rho_A),
\end{eqnarray}
where $\rho$ is the sum of the energy densities of the usual components in the
standard model, namely photons $\rho_\gamma$, neutrinos $\rho_\nu$,
matter $\rho_m$, and dark energy $\rho_{\rm DE}$. (In the following
discussion, we neglect the effects of matter since
nucleosynthesis takes place in radiation dominated era.)

The shear is sourced by the skewness according to the
equation~\cite{ellipsoidal6,ellipsoidal7}
\begin{equation}
\label{EvolShear} (H\Sigma)^{\cdot} + 3H^2\Sigma =
\frac{8\pi G}{3} \, (\rho_\nu \delta_\nu + \rho_A \delta_A),
\end{equation}
where $\delta_\nu$ is the neutrino skewness that takes care of
effects of anisotropic distribution of neutrinos. It depends on the shear
and its form will be discussed later.

Inflation generally causes an isotropization of the Universe:
any cosmic shear present before and/or during inflation will be
reduced to a vanishingly small value after inflation (see discussion below).
Nevertheless, if a source of anisotropy is present
after inflation (e.g. an anisotropic fluid), then the cosmic shear can grow and
be different from zero at the time of decoupling. If this is the case, planar cosmic symmetry
induces a quadrupole term in the CMB radiation which adds to that caused by the inflation-produced
gravitational potential at the last scattering surface.  If the planar-metric induced quadrupole
is comparable to the inflation-produced one, the overall quadrupole power may match the
``anomalously low'' value of the observed quadrupole~\cite{,ellipsoidal1,ellipsoidal2}.
The capability to solve the CMB quadrupole problem is the main
attractive feature of the ellipsoidal universe model.


\section{\normalsize{III. Helium-4 Synthesis}}
\renewcommand{\thesection}{\arabic{section}}

The mass fraction of $^4$He produced by
standard primordial nucleosynthesis at the cosmic time
$t_{\rm nuc}^{(0)} \simeq 300 \,$s
-- corresponding to a temperature of $T_{\rm nuc} \simeq 0.07 \MeV$ --,
is~\cite{Kolb}
\begin{equation}
\label{Y0}
Y^{(0)} \simeq \frac{2(n/p)_{\rm nuc}^{(0)}}{1 + (n/p)_{\rm nuc}^{(0)}} \simeq 0.25,
\end{equation}
where
\begin{equation}
\label{np0}
(n/p)_{\rm nuc}^{(0)} \simeq
e^{-Q/T_f^{(0)}} e^{-t_{\rm nuc}^{(0)}/\tau_n} \simeq 1/7
\end{equation}
is the neutron-to-proton number density ratio.
[We indicate quantities in the standard isotropic cosmological model
with an index ``(0)''.]
The first exponential factor in Eq.~(\ref{np0}), with $Q \simeq 1.3\MeV$
being the neutron-proton mass difference, is the neutron-proton number density ratio
at the time of freeze-out, namely when the expansion rate of the Universe, $H^{(0)}$,
equals the rate for the weak interactions, $\Gamma \sim G_F^2 T^5$,
that interconvert neutrons and protons ($G_F$ is the Fermi constant).
This happens at a temperature of about $T_f^{(0)} \simeq 0.8 \MeV$~\cite{Kolb}.
Due to ``deuterium bottleneck''~\cite{Kolb} the production of $^4$He is
delayed until the Universe has cooled to the temperature $T_{\rm nuc}$.
In this time lag, neutrons decay reducing their relative abundance and,
in turn, that of $^4$He. This gives the second exponential term in Eq.~(\ref{np0}), where
$\tau_n = 885\,$s is the mean neutron lifetime.

In ellipsoidal universe, both the freeze-out temperature and the time of
nucleosynthesis are modified, and so is $^4$He abundance.

Astrophysical observations fix the value of $^4$He abundance as
$Y^{(0)} \simeq 0.25 \pm 0.01$~\cite{Olive}. (See~\cite{PDB} and references
therein for more recent estimates of $Y^{(0)}$ which are, however, all
consistent with that quoted here. This can be considered as the most
conservative estimate of $Y^{(0)}$ since it possesses the largest uncertainty.)
Therefore, to be consistent with experimental data, we must require that the maximum variation
of $^4$He abundance (with respect to the isotropic case) is below the $4\%$.

A general expression for the freeze-out temperature in ellipsoidal universe
is easily obtained from Eq.~(\ref{H}) if one assumes that the energy content of the
anisotropic fluid is subdominant with respect to that of radiation:
\begin{equation}
\label{Tfreeze}
T_f = \frac{T_f^{(0)}}{(1-\Sigma^2_f)^{1/6}} \, ,
\end{equation}
where $\Sigma_f$ is the shear at the time of freeze-out and we used the fact that
$\rho = (\pi^2/30) g_* T^4$, with $g_*$ the total number of effectively massless degree of
freedom~\cite{Kolb}. In the following we simply assume $g_* = 3.36$ during nucleosynthesis
(even if $g_*$ increases from that value to $10.75$ near $T_f$).

The time when $^4$He is produced is easily found by integrating the Friedmann equation
with respect to time:
\begin{equation}
\label{tnuc}
t_{\rm nuc} = \frac{3\sqrt{5} \, m_{\rm Pl}}{2 \pi^{3/2} g_*^{1/2}}
\int_{T_{\rm nuc}}^\infty \frac{dT}{T^3} \, (1-\Sigma^2)^{1/2},
\end{equation}
where $m_{\rm Pl}$ is the Planck mass and we used the fact that $A \propto T^{-1}$.

Due to positivity of the energy and looking at the Friedmann equation we see that
the shear is bounded in the interval $[-1,1]$. This implies, using Eqs.~(\ref{Tfreeze})
and (\ref{tnuc}), that $T_f \geq T_f^{(0)}$ and $t_{\rm nuc} \leq t_{\rm nuc}^{(0)}$.
Since the $^4$He abundance, $Y$, is given by Eqs.~(\ref{Y0}) and (\ref{np0}) with
$T_f^{(0)}$ and $t_{\rm nuc}^{(0)}$ replaced by $T_f$ and $t_{\rm nuc}$, we conclude
that in ellipsoidal universe there is an
overproduction of $^4$He with respect to the isotropic case,
whatever is the nature of the anisotropic source.

In order to calculate this positive variation of $^4$He mass fraction, one has to specify
the anisotropic source so as to integrate Eq.~(\ref{EvolShear}), find the shear
as a function of temperature, and obtain $t_{\rm nuc}$ from Eq.~(\ref{tnuc}).

This can be done analytically only in the case where the effects of the skewness
are neglected ($\delta_A = \delta_\nu = 0$). Indeed, the case $\delta_A = 0$
is that studied numerically in the literature taking into account both the full set of nuclear
reactions leading to the production of light elements and the effects of anisotropic
distribution of neutrinos~\cite{Matzner1,Matzner2}. The effects of having $\delta_\nu \neq 0$
are studied below, but we will show that they are negligible (at least for small values of the shear).
\\
Introducing the anisotropy parameter $B = \Sigma^2/(1-\Sigma^2)$, the shear
equation~(\ref{EvolShear}) gives $B \propto T^2$, so we can easily
solve Eq.~(\ref{Tfreeze}) for the freeze-out temperature, and perform the integral in Eq.~(\ref{tnuc})
to get the time of nucleosynthesis:
\begin{eqnarray}
\label{TfreezeB}
&& T_f = T_f^{(0)} f_1[B(T_f^{(0)})] \, ,
\\
\label{tnucdelta=0}
&& t_{\rm nuc} = t_{\rm nuc}^{(0)} \, f_2[B(T_{\rm nuc})] \, ,
\end{eqnarray}
where
\begin{eqnarray}
\label{f1}
&& \!\!\!\!\!\!\!\!\! f_1[x] =
\sqrt{\frac{2 \times 3^{1\!/3} + 2^{1\!/3}(9+\sqrt{81-12x^3})^{2\!/3}}{6^{2\!/3}(9+\sqrt{81-12x^3})^{1\!/3}}} \, , \\
\label{f2}
&& \!\!\!\!\!\!\!\!\! f_2[x] = \sqrt{1+x} - x \, \mbox{arccosh} \sqrt{x} \, .
\end{eqnarray}
In Fig.~1, we plot the relative increase of $^4$He abundance
(with respect to the isotropic case) as a function of $B_0$,
namely the anisotropy parameter evaluated at the reference temperature
$T_0 = 50 \times 10^9 \K$, well before the nucleosynthesis starts.
($B$, and then $B_0$, are the same quantity introduced in~\cite{Matzner2}.)
The asymptotic expansions of such an increase, for small and large
values of $B_0$, are:
\begin{equation}
\label{Yapprox} \frac{Y-Y^{(0)}}{Y^{(0)}} =
    \left\{ \begin{array}{ll}
        f_3 B_0,  & \;\;\; B_0 \rightarrow 0, \\
        f_4,               & \;\;\; B_0 \rightarrow \infty,
    \end{array}
    \right.
\end{equation}
where
\begin{eqnarray}
\label{f3}
&& f_3 = \frac{(2-Y^{(0)}) \,Q \, T_f^{(0)}}{12\,T_0^2} + \mathcal{O}(\ln \! B_0) \simeq 0.01, \\
\label{f4}
&& f_4 = \frac{1-Y^{(0)}}{Y^{(0)}} \simeq 3.
\end{eqnarray}
The first term in the right hand side of Eq.~(\ref{f3})
takes into account the rise of the freeze-out temperature in anisotropic universe,
while the logarithmic term takes care of the reduction of time lag between the freeze-out and
the end of nucleosynthesis, and is negligible with respect to the first one.


\begin{figure}[t]
\begin{center}
\includegraphics[clip,width=0.47\textwidth]{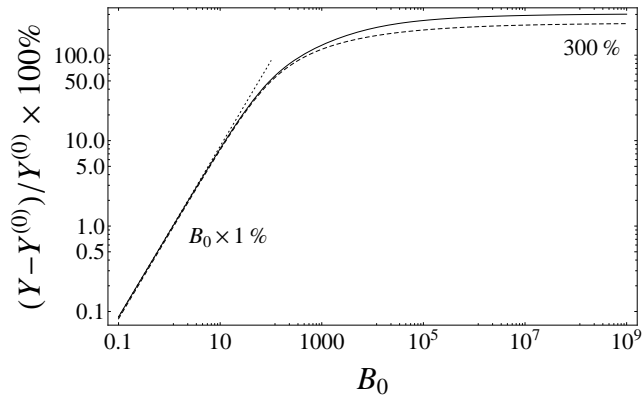}
\caption{Relative increase of $^4$He abundance in ellipsoidal universe with
zero skewness $\delta_A$ and neglecting neutrino anisotropy effects (continuous line)
as a function of the anisotropy parameter
$B_0 = B(T=50 \times 10^9 \K)$, where $B = \Sigma^2/(1-\Sigma^2)$.
The dotted line is the asymptotic expansion $B_0 \times 1\%$, while $300\%$ is
the limiting value for $B_0 \rightarrow \infty$. The dashed line is the relative
increase of $^4$He abundance in the same cosmological model but assuming no variation
on the time of nucleosynthesis, $t_{\rm nuc} = t_{\rm nuc}^{(0)}$.}
\end{center}
\end{figure}


The numerical analysis of~\cite{Matzner2} shows an increase of light element abundances.
In particular, the relative increase found for $^4$He is linear for moderate
values of $B_0$ ($B_0 \lesssim 10$) and is about $B_0 \times 3\%$.
Therefore, our oversimplified analytical model confirms qualitatively
(and to some extent also quantitatively) the numerical results of~\cite{Matzner2}.
\footnote{It is worth noticing that our analysis needs to be modified for very large values
of the anisotropy parameter $B$ since, as pointed out in~\cite{Barrow3}, the equilibrium of
weak interactions can be broken by very high levels of anisotropic expansion. However,
our forthcoming results will rely just on the part of Fig.~1 that corresponds to
moderate values of the anisotropy parameter (namely $B_0 \lesssim 10$), whose validity
has been already confirmed numerically in~\cite{Matzner2}.}

The dashed line in Fig.~1 is the relative increase of $^4$He abundance
assuming no variation on the time of nucleosynthesis, $t_{\rm nuc} = t_{\rm nuc}^{(0)}$.
As it is clear from the figure, the time delay effect due to cosmic anisotropy,
$t_{\rm nuc} \leq t_{\rm nuc}^{(0)}$, causes appreciable effects only for large shears which
are, however, unrealistic because of the large excess of $^4$He.

We note that if we just replace $T_f^{(0)}$ with $T_f$ and leave $t_{\rm nuc}^{(0)}$ in
Eqs.~(\ref{Y0}) and (\ref{np0}), we obtain a lower limit on $Y$. Imposing that the mass
fraction of $^4$He is less than $4\%$ with respect to the standard isotropic case, we
obtain a conservative, but model-independent limit (not depending on $\delta_A$ and $w_A$)
on the level of cosmic anisotropy at the time of freeze-out:
\begin{equation}
\label{limit}
|\Sigma(T_f)| \lesssim 0.4.
\end{equation}
It is straightforward to show that the above limit is in agreement with the
limit obtained by translating the current bound
on the total number of effectively massless degree of freedom at time of freeze-out.
Indeed, assuming as before that the energy content of the
anisotropic fluid is subdominant with respect to that of radiation,
we can rewrite Eq.~(\ref{H}) as the usual Friedmann equation
$H^2 = (8\pi G/3) \rho$ where now $\rho = (\pi^2/30) g_{*,\rm eff} \, T^4$
with
\begin{equation}
\label{R1}
g_{*,\rm eff} = \frac{g_*}{1-\Sigma^2} \, .
\end{equation}
Therefore, the effect of having a nonzero shear at the time of freeze-out
can be regarded as a change in the total number of effectively massless degree of freedom,
which is usually parameterized by the effective number of neutrino species, $N_\nu$, as~\cite{Kolb}
\begin{equation}
\label{R2}
g_{*,\rm eff}  = \frac{11}{2} + \frac74 N_\nu \left(\frac{4}{11}\right)^{\!4/3} \! .
\end{equation}
(The standard value $g_* \simeq 10.75$ near $T_f$ corresponds to take $N_\nu = 3$ in the above equation.)
Using the current bound $N_\nu = 3.2 \pm 1.2$ ($95\% \; \mbox{C.L.}$)~\cite{PDB} on the
effective number of neutrino species at the time of freeze-out, and comparing
Eqs.~(\ref{R1}) and (\ref{R2}), we get $|\Sigma(T_f)| = 0.11 \pm 0.34$ ($95\% \; \mbox{C.L.}$),
where we used the Gauss error propagation law to propagate the uncertainty on $N_\nu$ to $|\Sigma(T_f)|$.
So, we obtain the upper bound $|\Sigma(T_f)| \lesssim 0.45$, which is compatible with Eq.~(\ref{limit}).

The effects of anisotropic distribution of neutrinos can be described as follows.
For temperature greater than  $T_f^{(\nu)} = \mathcal{O}(\MeV)$, neutrinos are strongly coupled
to primordial plasma, so their distribution is isotropic and no neutrino skewness
results. Below a temperature slightly lower than $T_f^{(\nu)}$, instead,
neutrinos begin to free-stream and generate a skewness
\begin{equation}
\label{deltanu}
\delta_\nu(T) = \frac{8}{5} \int_{T_f}^{T} \frac{dT'}{T'} \, \Sigma(T').
\end{equation}
Here, for the sake of simplicity, we assumed an instantaneous neutrino decoupling
at $T_f^{(\nu)} \simeq T_f$, so neutrino free-streaming affects only the time
when $^4$He is produced, namely $t_{\rm nuc}$.
The above result~(\ref{deltanu}) is valid for small values of the shear ($|\Sigma| \ll 1$)
and can be obtained from~\cite{Matzner1} taking the limit, in the Misner's anisotropy potential,
of large collision time ($t_c \rightarrow \infty$) for the
typical reactions of neutrinos with plasma.

Using~(\ref{deltanu}) in Eq.~(\ref{EvolShear}), we get for $T \leq T_f$
\begin{equation}
\label{Sigmanu}
\Sigma = \Sigma_f \left(\frac{T}{T_f}\right)^{\!\!1/2} \!
\left \{ \!\cos \! \left[c_\nu \ln \! \frac{T}{T_f}\right] +
\frac{1}{c_\nu} \sin\! \left[c_\nu \ln \! \frac{T}{T_f} \right] \!\right \} \! ,
\end{equation}
where
$c_\nu = \sqrt{8\Omega_\nu/5 -1/4}$
with $\Omega_\nu = \rho_\nu/\rho_{\rm cr}$ being the neutrino energy density parameter and
$\rho_{\rm cr} = 3H^2/8\pi G$ the critical density.
The neutrino energy density parameter $\Omega_\nu$ is constant during radiation
era and, assuming three neutrino massless species, equal to about 0.4 after neutrino
decoupling~\cite{Kolb}, so $c_\nu \simeq 0.6$. The shear is an oscillating function of
time with a damping factor proportional to $T^{-1/2}$. This leads to very tiny
variations of $t_{\rm nuc}$ with respect to the case $\delta_\nu = 0$,
and gives rise to a small increase of $(Y-Y^{(0)})/Y^{(0)} \times 100\%$,
which is below the $0.15\%$ for $0 \leq B_0 \lesssim 1$
(corresponding to $\Sigma \lesssim 0.3$).

Since the absolute value of the shear at the time of freeze-out must be significatively
smaller than one [see Eq.~(\ref{limit})], we can now consider a simplified
(but more realistic) model where the shear is a small quantity during radiation era and the
effects of both neutrino skewness $\delta_\nu$ and external anisotropic sources $\delta_A$
are taken into account.

For small shears and $\delta_A$ constant, the energy density of an anisotropic
fluid evolves as in the case of isotropic universe,
$\rho_A \propto A^{-3(1+w_A)}$~\cite{ellipsoidal6,ellipsoidal7}, and the shear
equation~(\ref{EvolShear}) can be solved to give in radiation era and before neutrino decoupling,
\begin{equation}
\label{SigmaSol}
\Sigma = \frac{\mbox{constant}}{A} + \frac{\delta_A}{2-3w_A} \, \frac{\Omega_{A,0}}{\Omega_{r,0}} \, A^{1-3w_A},
\end{equation}
where we assumed that the energy density of the anisotropic component is
small with respect to that of radiation. This is the same as assuming
$w_A < 1/3$, or $\delta_A \Omega_{A,0} \ll \Omega_{r,0}$ if $w_A = 1/3$.
Here, $\Omega_{A,0}$ and $\Omega_{r,0}$ are the present energy density
parameters of anisotropic fluid and radiation, respectively.
\footnote{For $w_A=1/3$, Eq.~(\ref{SigmaSol}) is correct up to a logarithmic term. Indeed,
as shown in~\cite{Barrow1}, the last term in the right hand side of Eq.~(\ref{SigmaSol})
should be divided, in this case, by $1 + 2 \delta_A^2 (\Omega_{A,0}/\Omega_{r,0}) \ln (A/A_{\rm end})$,
where $A_{\rm end}$ is the expansion parameter at the end of inflation.
However, the inclusion of this term modifies
Eq.~(\ref{SigmaSol}) only to second order in the small quantity $\delta_A \Omega_{A,0}/\Omega_{r,0}$.
Therefore, for simplicity, we neglect this term in the following.}

We can fix the integration constant in Eq.~(\ref{SigmaSol}) by
evaluating the shear at early times, for example at the end of
inflation, $A = A_{\rm end} \ll 1$.
If $w_A < 1/3$ we get $\mbox{constant} \simeq A_{\rm end} \Sigma_{\rm end}$,
where $\Sigma_{\rm end} = \Sigma(A_{\rm end})$, while for $w_A = 1/3$
we have $\mbox{constant} \simeq A_{\rm end} (\Sigma_{\rm end} -\delta_A \Omega_{A,0}/\Omega_{r,0})$.
As shown in~\cite{Barrow1}, any cosmic anisotropy is washed out (exponentially)
during (de Sitter) inflation (for $w_A > -1$), so that anisotropy can develop just
at the end of inflation starting from a vanishingly small value.
\footnote{In de Sitter inflation, subdominant anisotropic fluids are such that $w_A > -1$, or
$\delta_A \Omega_{A,0} \ll 1$ if $w_A = -1$. For such fluids and in the limit of small shears, it is easy
to see that $\Sigma_{\rm end} \simeq \Sigma_i \, e^{-3N}$ if $w_A > 0$ and
$\Sigma_{\rm end} \simeq -(\delta_A/3w_A) \, \Omega_{A,0} \, e^{-3(1+w_A)N}$ if $-1 \leq w_A < 0$ and
$A \gg A_i$. Here, $\Sigma_i$ is the shear at the beginning of inflation at $A = A_i$,
and $N \gtrsim 60$ the number of $e$-folds of inflation since inflation began~\cite{Kolb}.}
This means that $\Sigma_{\rm end} \simeq 0$, so that we can neglect the first term in
the right hand side of Eq.~(\ref{SigmaSol}) for $A \gg A_{\rm end}$.

Let us assume for the moment that the effects of neutrino free-streaming are negligible.
In this case, the above solution~(\ref{SigmaSol}) is valid throughout nucleosynthesis
and, since $\Sigma \ll 1$, we conclude that no appreciable changes on $^4$He production
occur with respect to the isotropic case. We can now
check the validity of the assumption of neglecting neutrino anisotropy.
By inserting Eq.~(\ref{SigmaSol}) in Eq.~(\ref{deltanu}) we find that the ratio of the anisotropy
sources in Eq.~(\ref{EvolShear}) is, for $T \leq T_f$ and $w_A \neq 1/3$:
\begin{equation}
\label{ratiodelta}
\frac{\rho_\nu \delta_\nu}{\rho_A \delta_A}  =
-\frac{8}{5} \, \frac{\Omega_{\nu,0}}{\Omega_{r,0}} \, \frac{1-(T/T_f)^{1-3w_A}}{(1-3w_A)(2-3w_A)} \, ,
\end{equation}
where $\Omega_{\nu,0}$ is the present neutrino energy density parameter. Assuming three
neutrino massless species, we have $\Omega_{\nu,0}/\Omega_{r,0} \simeq 0.4$~\cite{Kolb}.
For the cosmologically interesting cases of anisotropic dark energy ($w_A \simeq -1$),
a cosmic domain wall ($w_A = -2/3$), and a cosmic string ($w_A = -1/3$), 
the absolute value of the ratio~(\ref{ratiodelta}) is maximum at $T = T_{\rm nuc}$
and is much smaller than one
(0.03, 0.05, and 0.11, respectively, assuming $T_f \simeq T_f^{(0)}$), 
and this justifies our previous assumption.

The case $w_A = 1/3$, namely an anisotropic component of radiation type, has to be analyzed separately.
In this case, the shear equation~(\ref{EvolShear}) can be solved and gives, for $T \leq T_f$,
Eq.~(\ref{Sigmanu}) with the factor $1/c_\nu$ multiplying the sine function replaced by
$c_A/c_\nu$, where $c_A = \delta_A \Omega_A/\Sigma_f - 1/2$.
Using Eq.~(\ref{SigmaSol}) evaluated at $T=T_f$, we find $c_A = \Omega_{r,0} \Omega_A/\Omega_{A,0}-1/2$.
Since both $\Omega_A$ and $\Omega_r$ are constant in radiation era and scale as
$T$ in matter era, we get $c_A = \Omega_r-1/2$ in radiation era, where
$\Omega_r = 1 - \Omega_\nu \simeq 0.6$ after neutrino decoupling. Therefore $c_A \simeq 0.1$.
Since neglecting neutrino anisotropy we found that $\Sigma$ is constant (up to a logarithmic correction)
in radiation era [see Eq.~(\ref{SigmaSol})] and does not affect Helium-4 synthesis,
we conclude that also in the case $\delta_\nu \neq 0$, where the shear is an oscillating function of
time with a damping factor proportional to $T^{-1/2}$,
no appreciable changes on $^4$He production occur with respect to the isotropic case.

Before concluding, let us include in our analysis a component of free-streaming gravitons,
for inflation generally predicts gravitational waves, namely tensor fluctuations, which
are not in thermal equilibrium below the Planck scale and then can be considered
as free-streaming radiation from inflation until today.
Gravitational waves introduce the extra term $(8\pi G/3) \rho_{\rm GW} \delta_{\rm GW}$
on the right-hand-side of Eq.~(\ref{EvolShear}), where $\rho_{\rm GW}$ and $\delta_{\rm GW}$
are the graviton energy density and skewness, respectively.
The energy density associated to this background of gravitational waves is typically very small:
$\Omega_{\rm GW,0}/\Omega_{r,0} \lesssim 10^{-8}$ for modes that cross inside the horizon while
the Universe is radiation dominated~\cite{Kolb},
where $\Omega_{\rm GW,0}$ is the present energy density parameter of gravitons.
The graviton skewness is given, after inflation,
by an expression similar to Eq.~(\ref{deltanu})
\begin{equation}
\label{R3}
\delta_{\rm GW}(T) = \frac{8}{5} \int_{T_{\rm RH}}^{T} \frac{dT'}{T'} \, \Sigma(T'),
\end{equation}
where $T_{\rm RH}$ is the so-called reheating temperature, that is the
temperature of the cosmic plasma at the beginning of the radiation
era. (Here and in the following we assume that the reheating phase,
during which the energy of the inflaton is converted into ordinary
matter is ``instantaneous'' so that, after inflation, the universe
enters directly into the radiation era.)
Now it is easy to show that, due to the smallness of $\Omega_{\rm GW,0}$, the effect
of gravitational waves in the evolution of the shear is completely negligible. In fact,
proceeding as we did in obtaining Eq.~(\ref{ratiodelta}), we can verify that the ratio
$|\rho_{\rm GW} \delta_{\rm GW}/\rho_A \delta_A|$ is much smaller than unity. Indeed,
for $T \leq T_{\rm RH}$ and $w_A \neq 1/3$, it is given by the right hand side of
Eq.~(\ref{ratiodelta}) with $\Omega_{\nu,0}$ and $T_f$
replaced by $\Omega_{\rm GW,0}$ and $T_{\rm RH}$, respectively. Therefore we have
$|\rho_{\rm GW} \delta_{\rm GW}/\rho_A \delta_A| \sim \Omega_{\rm GW,0}/\Omega_{r,0} \ll 1$.
For $T \leq T_{\rm RH}$ and $w_A = 1/3$ we get, instead,
\begin{equation}
\label{R5}
\frac{\rho_{\rm GW} \delta_{\rm GW}}{\rho_A \delta_A}  =
-\frac{8}{5} \, \frac{\Omega_{\rm GW,0}}{\Omega_{r,0}} \, \ln (T_{\rm RH}/T).
\end{equation}
The absolute value of the above ratio is maximum for $T = T_{\rm nuc}$ and for
the largest allowed value of $T_{\rm RH}$, $T_{\rm RH} \simeq 10^{17} \GeV$~\cite{Kolb}.
Also in this case it is much smaller than unity:
$|\rho_{\rm GW} \delta_{\rm GW}/\rho_A \delta_A| \simeq 78 \, \Omega_{\rm GW,0}/\Omega_{r,0} \ll 1$.

Let us conclude by observing that for a uniform magnetic field $w_A = 1/3$ and $\delta_A = 2$.
Therefore, the above results show that uniform magnetic fields created at inflation
and whose energy density is small with respect to that of radiation
do not affect nucleosynthesis.
However, in the presence of an external uniform magnetic field, nucleosynthesis is affected,
other than by the effect of anisotropization of the Universe due to a nonvanishing shear,
also by the increase of weak reaction rates, of the expansion rate of the Universe,
and of the electron density~\cite{Cheng}. Taking into account all these effects, but not
the effect here studied of nonvanishing $\Sigma$,
the authors of~\cite{Cheng} found that observations of light elements are
compatible with a magnetic field energy density lower than $\rho_\mathcal{B} \lesssim 0.28 \rho_\nu$,
where $\rho_\mathcal{B} = \mathcal{B}^2/2$ is the magnetic energy density associated
to a uniform magnetic field of intensity $\mathcal{B}$.
Their analysis is correct as long as the effect of the shear can be neglected
which means, in light of the previous discussion, that the magnetic field must be
a subdominant component of the Universe during nucleosynthesis.
This is indeed the case, since the subdominance condition
for a uniform magnetic field, $2 \Omega_{\mathcal{B},0} \ll \Omega_{r,0}$,
translates to
\begin{equation}
\label{rhoB}
\rho_\mathcal{B} \ll \rho_\mathcal{B}^{\rm max} = \frac{\rho_\nu }{2\Omega_\nu} \, \simeq 1.25 \rho_\nu
\end{equation}
after neutrino decoupling, a limit about 5 times greater than that allowed
by the analysis of~\cite{Cheng}.

It is worth noticing that the above limit on the intensity of a cosmological
magnetic field is much less stringent than that coming from the analysis of the
CMB radiation~\cite{Barrow2,ellipsoidal1,ellipsoidal2,ellipsoidal7,Bunn}
which is at least two order of magnitude stronger. This agrees with Barrow's
result~\cite{Barrow1} that anisotropic fluids that create temperature anisotropies
compatible with CMB spectrum do not have a significant effect on the primordial
synthesis of $^4$He.


\section{\normalsize{IV. Conclusions}}
\renewcommand{\thesection}{\arabic{section}}

In this paper, we have analyzed the effects caused by cosmic
anisotropy on the primordial production of $^4$He. We worked in the context of a
cosmological model of Bianchi type I where the anisotropy of spatial geometry, the
shear $\Sigma$, is generated by a fluid with anisotropic equation of state.

We found that in such an anisotropic universe there is an overproduction of $^4$He
with respect to the standard isotropic case. Imposing that the relative increase of
$^4$He abundance is below the $4\%$ to be consistent with observational data, we
constrained the absolute value of the shear to be less than $0.4$ at the time of
freeze-out. This limit does not depend on the equation(s) of state of the anisotropic
fluid and has been obtained assuming that the energy density of the
anisotropic fluid is small compared to that of radiation.

Moreover, we showed  that anisotropic fluids generated at inflation,
such as dark energy with anisotropic equation of state and a uniform
magnetic field, create anisotropies much smaller than the above limit
if their energy densities are subdominant with respect to that of the
Universe during inflation and radiation era. In particular, the existence
of a uniform magnetic field at the time of nucleosynthesis is compatible
with astrophysical data if its energy density is much smaller than about
$1.25$ times the energy density of neutrinos.


\begin{acknowledgments}
We would like to thank P. Cea and J. D. Barrow for very helpful discussions.
\end{acknowledgments}



\begin{thebibliography}{99}

\bibitem{Weinberg}         S. Weinberg,
                           {\it Cosmology}
                           (Oxford University Press, New York, New York, 2008).

\bibitem{ellipsoidal1}     L.~Campanelli, P.~Cea and L.~Tedesco,
                           Phys.\ Rev.\ Lett.\ {\bf 97}, 131302 (2006);
                           {\bf 97}, 209903(E) (2006).

\bibitem{ellipsoidal2}     L.~Campanelli, P.~Cea and L.~Tedesco,
                           Phys.\ Rev.\ D {\bf 76}, 063007 (2007).

\bibitem{Barrow1}          J.~D.~Barrow,
                           Phys.\ Rev.\ D {\bf 55}, 7451 (1997).

\bibitem{Koivisto-Mota}    T.~Koivisto and D.~F.~Mota,
                           Astrophys.\ J.\ {\bf 679}, 1 (2008).

\bibitem{R1}               S.~Hawking and J.~R.~Tayler,
                           Nature\ {\bf 209}, 1278  (1966).

\bibitem{R2}               K.~S.~Thorne,
                           Astrophys.\ J.\ {\bf 148}, 51  (1967).

\bibitem{R3}               R.~F.~Carswell,
                           Mon.\ Not.\ R.\ Astron.\ Soc.\ {\bf 144}, 279 (1969).

\bibitem{R4}               J.~D.~Barrow,
                           Mon.\ Not.\ R.\ Astron.\ Soc.\ {\bf 175}, 359 (1976);
                           {\bf 211}, 221 (1984).

\bibitem{ellipsoidal4}     L.~Campanelli,
                           Phys.\ Rev.\ D {\bf 80}, 063006 (2009).

\bibitem{ellipsoidal3}     P.~Cea,
                           arXiv:astro-ph/0702293.

\bibitem{ellipsoidal5}     P.~Cea, 
                           Mon.\ Not.\ Roy.\ Astron.\ Soc.\ {\bf 406}, 586 (2010).

\bibitem{ellipsoidal6}     L.~Campanelli, P.~Cea, G.~L.~Fogli and L.~Tedesco,
                           Int.\ J.\ Mod.\ Phys.\ D {\bf 20}, 1153 (2011).

\bibitem{ellipsoidal7}     L.~Campanelli, P.~Cea, G.L.~Fogli, and L.~Tedesco,
                           Mod.\ Phys.\ Lett.\ A {\bf 26}, 1169 (2011).

\bibitem{ellipsoidal8}     L.~Campanelli, P.~Cea, G.~L.~Fogli and A.~Marrone,
                           Phys.\ Rev.\ D {\bf 83}, 103503 (2011).

\bibitem{Taub}             A.~H.~Taub,
                           Annals Math.\ {\bf 53}, 472 (1951).

\bibitem{Kolb}             E.~W.~Kolb and M.~S.~Turner,
                           {\it The Early Universe}
                           (Addison-Wesley, Redwood City, California, 1990).

\bibitem{Olive}            K.~A.~Olive and E.~D.~Skillman,
                           Astrophys.\ J.\ {\bf 617}, 29 (2004).

\bibitem{PDB}              C.~Amsler et al. [Particle Data Group],
                           Phys.\ Lett.\ {\bf B667}, 1 (2008),
                           and 2009 partial update for the 2010 edition.

\bibitem{Matzner1}         T.~Rothman and R.~Matzner,
                           Phys.\ Rev.\ D {\bf 30}, 1649 (1984).

\bibitem{Matzner2}         H.~Kurki-Suonio and R.~Matzner,
                           Phys.\ Rev.\ D {\bf 31}, 1811 (1985).

\bibitem{Barrow3}          J.~D.~Barrow,
                           Mon.\ Not.\ R.\ Astron.\ Soc.\ {\bf 178}, 625 (1977).

\bibitem{Cheng}            B.~l.~Cheng, A.~V.~Olinto, D.~N.~Schramm and J.~W.~Truran,
                           Phys.\ Rev.\ D {\bf 54}, 4714 (1996).

\bibitem{Barrow2}          J.~D.~Barrow, P.~G.~Ferreira and J.~Silk,
                           Phys.\ Rev.\ Lett.\ {\bf 78}, 3610 (1997).

\bibitem{Bunn}             E.~F.~Bunn, P.~Ferreira and J.~Silk,
                           Phys.\ Rev.\ Lett.\ {\bf 77}, 2883  (1996).

\end{thebibliography}
\end{document}